\pdfoutput=1
\documentclass[a4paper,11pt]{article}

\usepackage{jheppub,undertilde}

\newcommand{\beq}{\begin{eqnarray}}
\newcommand{\eeq}{\end{eqnarray}}
\newcommand{\be}{\begin{equation}}
\newcommand{\ee}{\end{equation}}
\def\bsp#1\esp{\begin{split}#1\end{split}}

\preprint{CERN-PH-TH/2014-140, LYCEN 2014-11}
\title{Revisiting monotop production at the LHC}

\author[a]{Idir Boucheneb}
\author[a,b]{, Giacomo Cacciapaglia}
\author[a,b,1]{, Aldo Deandrea\note{also Institut Universitaire de France, 103 boulevard Saint-Michel, 75005 Paris, France}}
\author[c,d]{and Benjamin Fuks}

\affiliation[a]{Universit\'e de Lyon, F-69622 Lyon, France; Universit\'e Lyon 1,
   Villeurbanne}
\affiliation[b]{CNRS/IN2P3, UMR5822, Institut de Physique Nucl\'eaire de Lyon
   F-69622 Villeurbanne Cedex, France}
\affiliation[c]{CERN, PH-TH, CH-1211 Geneva 23, Switzerland}
\affiliation[d]{Institut Pluridisciplinaire Hubert Curien/D\'epartement
    Recherches Subatomiques, Universit\'e de Strasbourg/CNRS-IN2P3,
    23 rue du Loess, F-67037 Strasbourg, France}

\emailAdd{g.cacciapaglia@ipnl.in2p3.fr}
\emailAdd{deandrea@ipnl.in2p3.fr}
\emailAdd{benjamin.fuks@cern.ch}

\abstract{
Scenarios of new physics where a single top quark can be produced in association with large missing energy (monotop) have
been recently studied both from the theoretical point of view and by experimental collaborations.
We revisit the originally proposed monotop setup by embedding the effective couplings of the
top quark in an SU(2)$_L$ invariant
formalism. We show that minimality selects one model for each of the possible
production mechanisms: a scalar field coupling to a right-handed top quark and an invisible fermion when the monotop
system is resonantly produced, and a vector field mediating the interactions
of a dark sector to right-handed quarks for
the non-resonant production mode.
We study in detail constraints on the second class of scenarios, originating from contributions to
standard single top processes when the mediator is lighter than the top quark
and from the dark matter relic abundance when the mediator is heavier than the top quark.
}


\begin{document}
\maketitle
\flushbottom

\section{Introduction}
\label{sec:intro}
The first phase of the LHC experiments has given two important messages: a scalar resonance closely resembling the Standard Model Higgs boson
has been discovered, and new physics beyond the Standard Model has not been found. The latter result would imply that new states or effects beyond the Standard Model predictions may be much more difficult to spot at the LHC than we previously thought. In fact, very strong bounds have been posed on {\it easy--catch} models, like the
constrained version of the Minimal Supersymmetric Standard Model~\cite{atlas:susytwiki,cms:susytwiki}.

Many theorists have therefore recently turned their attention on a more signature-based strategy,
focusing on unusual final states which are difficult to detect or have not been considered yet by experimental collaborations.
One such final state that has been gaining popularity among
phenomenologists~\cite{Berger:1999zt,delAguila:1999ac,Berger:2000zk,Desai:2010sq,%
Davoudiasl:2011fj,Andrea:2011ws,Kamenik:2011nb,Dong:2011rh,Wang:2011uxa,Fuks:2012im,Kumar:2013jgb,Alvarez:2013jqa,Agram:2013wda,Fuks:2014uka,Ng:2014pqa} and
experimentalists~\cite{Aaltonen:2012ek,CMS:2014hba,Khachatryan:2014uma,Aad:2014wza}
is the monotop signature: a single top quark produced in association with a large amount of missing energy.
Although the production of this final state is very suppressed in the Standard Model,
it is however not easy to obtain this kind of events in realistic and complete models of new physics.
Two main production mechanisms can lead to a monotop state~\cite{Andrea:2011ws,Agram:2013wda}, arising either from
the resonant production of a coloured bosonic state which further decays into a
top quark plus an invisible neutral fermion; or via the production of a single top quark
in association with an invisible boson that has flavour-changing couplings to top and light quarks.
Examples of the first class of models include $R$-parity violating supersymmetry, where the
produced resonance is a top squark decaying into a top plus a long-lived
neutralino~\cite{Berger:1999zt,Berger:2000zk,Desai:2010sq,Fuks:2012im}.
The second class of models has
been described in scenarios of dark matter from a hidden sector that couples to the Standard Model
via flavour-violating couplings of a bosonic mediator~\cite{Davoudiasl:2010am,Davoudiasl:2011fj,Kamenik:2011nb,Alvarez:2013jqa}.
In general, monotop signatures can be however generated by other processes
involving, for instance, the $t$-channel exchange of a new particle, different spin
assigments for the new states or higher-spin tensors. Motivated by the
setups currently under study experimentally, this work is limited
to the case of a spin-0 or spin-1 state that can be either
exchanged in the $s$-channel (named ``resonant'' in the following) or produced in
association with a top quark via flavour-changing interactions
(named ``non-resonant'').

All such models can be described in terms of a simple Lagrangian~\cite{Andrea:2011ws}, which contains all the possible 
couplings giving rise to a monotop signal. A very
general analysis of this framework can be found in Ref.~\cite{Agram:2013wda}, the limiting
case of higher-dimensional operators has been discussed in Ref.~\cite{Dong:2011rh},
while monotop production via flavour-changing interactions of quarks with
an invisible $Z$-boson has been detailed in Ref.~\cite{delAguila:1999ac}.
Although this simple description has the advantage of being complete, it has the drawback of containing too many free parameters to be efficiently scanned by an experimental search.
Furthermore, the included couplings do not respect the symmetries of the Standard Model,
as they are intended to describe the model dynamics after the breaking of the electroweak symmetry.
In this way, this approach ignores other interactions needed to restore gauge invariance which can
give rise to new physics signals in different search channels, the latter possibly
implying stronger constraints on the parameters of the model than the monotop search itself.
In this work, we revisit the parametrisation originally proposed in
Ref.~\cite{Andrea:2011ws} by paying particular attention to the embedding of the
Lagrangian description into $SU(2)_L \times U(1)_Y$ invariant operators.
We therefore present a set of minimal effective Lagrangians, extending the Standard Model with gauge-invariant operators. Our
approach allows us to restrict the number of ``interesting'' scenarios, \textit{i.e.},
the cases where the monotop signal is genuinely the main signal of new physics to be expected at
the LHC. Equivalently, this reduces the number of free parameters to a manageable number.
Finally, we discuss in detail how the effective model could be completed in order to
guarantee that the missing energy particle produced in association with the top quark
is indeed either long-lived or decaying into invisible states.

The rest of this work is organized as follows. In Section~\ref{sec:modelbuilding},
we present how to construct gauge-invariant effective models for the monotop signal. 
We consider separately the resonant case, where the mediator is a scalar or vector boson, and the non resonant case where the top is produced in association with a scalar or vector via flavour-changing couplings. We also discuss how the effective Lagrangians can be matched to the simple monotop descriptions of Ref.~\cite{Andrea:2011ws}.
We then focus on non-resonant scenarios which turn out to be less
``standard'' and investigate, in Section~\ref{sec:nonrespheno}, the conditions under which the
invisible state is effectively
invisible, and other experimental observations, which can further constrain
the model. Our conclusions are presented in Section~\ref{sec:conclusions}.

\section{Gauge-invariant effective Lagrangians for monotop production}\label{sec:modelbuilding}

\subsection{Resonant monotop production}
\label{sec:lres}

In the first class of scenarios yielding the production of
a monotop system at colliders that we consider, the produced top quark recoils against
an invisible fermionic state $\chi$.
Being singly produced, the $\chi$ particle cannot be stable, thus it is either
long-lived or it decays into a pair of stable particles. In each case, it has to be
electrically-neutral and a
colour-singlet.
In resonant monotop production, both final-state particles arise from the decay of a heavy scalar $\varphi$ or vector $X$ field, lying
in the fundamental representation of $SU(3)_c$, that is produced in the $s$-channel
from the fusion of two down-type (anti-)quarks.
The parton-level process that we want to focus on is therefore
\beq
d d \to \varphi, X_\mu \to \bar{t} \chi\,.
\eeq
In order to understand how the scalar or vector mediators transform under the full
Stanard Model symmetries, it is crucial to define the chirality of the quarks it couples too.
In the following we will analyse separately the scalar and vector case in detail.

\subsubsection*{Spin-0 mediator}
The initial state consists of a pair of down-type quarks which form a spin-0 state. The related
Lorentz scalar fermionic bilinear reads $\bar{\psi} \psi$, which can be written as
$\bar{\psi}_L \psi_R + \bar{\psi}_R \psi_L$. This implies that the two quarks
have opposite chiralities. Recalling that the charge conjugate of the right-handed quark $d^C_R$
is left-handed while the one of the left-handed quark $d^C_L$ is right-handed, we can define two
independent initial states, one with explicit right-handed and one with left-handed chirality indices.
According to the Standard Model gauge symmetries $SU(3)_c$, $SU(2)_L$ and $U(1)_Y$, the two states transform as:
\be\bsp
  \bar{d}_R^{C\; i} d_R^j = ({\utilde{\bf \bar{3}}}, {\utilde{\bf 1}},  -2/3)\,; \qquad
  \bar{d}_L^{C\; i} d_L^j = ({\utilde{\bf \bar{3}}}, {\utilde{\bf 3}}, 1/3) \,.
\esp\ee
In our notation, an undertilde indicates a representation under a non-Abelian gauge symmetry, and the indices $i,j$ refer to flavour.
Since the diquark states are made of identical quarks and fermions are
anticommuting quantities, the corresponding wavefunctions need to be
antisymmetric under the exchange of the quark fields.
The exchange of the flavour indices is therefore forced to be antisymmetric too,
since the one of the spin and colour indices (which we do not explicit for
simplicity) are antisymmetric and the one of the triplet (adjoint) representation
of SU(2)$_L$ is symmetric (for the left-handed quark setup).
From the representations shown above, the right and left-handed quarks cannot couple to the same scalar and two different objects must thus be introduced,
\beq
\varphi_s = ({\utilde{\bf 3}}, {\utilde{\bf 1}},  2/3)\,; \qquad \varphi_t = ({\utilde{\bf 3}}, {\utilde{\bf 3}}, -1/3) \equiv
   \left( \begin{array}{c} \varphi_t^{2/3} \\ \varphi_t^{-1/3} \\ \varphi_t^{-4/3} \end{array} \right)\,,
\label{eq:scafields}\eeq
where the subscript $_s$ and $_t$ refer to singlet and triplet of SU(2)$_L$.
The operators containing the interactions needed for monotop production
can be written as
\beq
  \lambda_s\ \varphi_s\, \bar{d}_R^C d_R + \lambda_t\ \varphi_t \,\bar{q}_L^C q_L
  + {\rm h.c.} \ ,
\label{eq:lresGauge}\eeq
where $q_L$ is the left-handed doublet, and $\lambda_{s,t}$ are antisymmetric matrices in flavour space.
In the second term, it is the component of the triplet with electric charge $\pm 2/3$ that couples eventually to the top.

We can now repeat the same analysis for the final state. We first assume,
for minimality, that the invisible fermion $\chi$ is a singlet under
the Standard Model symmetries.
In this case, the final state system representation under the
Standard Model gauge group can be
\be\bsp
  \chi_R t_R = ({\utilde{\bf 3}}, {\utilde{\bf 1}},  2/3)\,; \qquad
  \chi_L t_L = ({\utilde{\bf 3}}, {\utilde{\bf 2}}, 1/6) \,.
\esp\ee
In order to allow for a $\chi$-coupling to a left-handed top quark,
we therefore need to introduce an extra scalar field $\varphi_d$, compared
to Eq.~\eqref{eq:scafields}, which transforms as a doublet of
$SU(2)_L$.
The operators relevant for monotop production can then be written as
\beq
y_s\ \varphi_s^\dagger \, \bar{\chi} t_R + y_d\ \varphi_d^\dagger \bar{\chi} q_L 
  + {\rm h.c.}
\eeq
The initial and final state can consequently only be connected via an
$SU(2)_L$-singlet field $\varphi_s$ that couples to right-handed quarks.
This scenario being minimal, it will therefore be considered in the rest of
this work. For completeness, let us mention that
non-minimal models with several additional scalars could also
be constructed. In these setups, the new scalar fields are allowed to mix
after electroweak symmetry breaking through couplings to the
Brout-Englert-Higgs field
$\phi_H = ({\utilde{\bf 1}}, {\utilde{\bf 2}},  1/2)$,
\beq
  \mu_t\ \phi_H^\dagger  \varphi_t^\dagger \varphi_d+
  \mu_d\ \varphi_s \varphi_d^\dagger \phi_H^\dagger + {\rm h.c.}
\eeq
The resulting mass
splitting is nevertheless constrained to be small by the perturbativity of the
couplings~\cite{Jack:1982sr} and corrections to the $S$ and $T$
parameters~\cite{Li:1992dt,Bhattacharyya:1993zy}.

We have also imposed that the $\chi$-field has the same quantum numbers
as a right-handed neutrino, so that it could potentially mix
with neutrinos. This mixing is however strongly constrained
by proton-decay processes like $p \to \pi^+/K^+ \nu$. In this case,
the contribution of the box-diagram-induced subprocess
$d u \to \bar{d}/\bar{s} \nu$ (through a $\varphi_s$ and $W^+$
exchange) has indeed to be
maintained small. There is however a way to evade the bound by
preventing $\chi$ from mixing with neutrinos (by assigning it, \textit{e.g.}, a
baryon number), unless $\chi$ is lighter than the proton.

We have considered so far models where the invisible fermion
$\chi$ is a Standard Model gauge singlet. Another option would be
to assume that $\chi$ is the neutral component of a non-trivial
$SU(2)_L$ multiplet. For instance, one could choose (with $\sigma_2$ being the
second Pauli matrix)
\be\bsp
  \chi_d = ({\utilde{\bf 1}}, {\utilde{\bf 2}},  1/2) \equiv
    \begin{pmatrix}  \chi^+ \\ \chi^0 \end{pmatrix}
    \qquad &\ \Rightarrow \qquad
    \chi_L t_L = ({\utilde{\bf 3}}, {\utilde{\bf 1}}, 2/3) \ ,\\
  \chi^\prime_d = i \sigma_2 \chi_d^\ast =  ({\utilde{\bf 1}}, {\utilde{\bf 2}},  -1/2) \equiv
    \begin{pmatrix} \chi^0 \\ \chi^- \end{pmatrix}
    \qquad &\ \Rightarrow \qquad
    \chi_L t_L = ({\utilde{\bf 3}}, {\utilde{\bf 3}}, -1/3) \ ,\\
  \chi_t = ({\utilde{\bf 1}}, {\utilde{\bf 3}}, -1) \equiv
    \begin{pmatrix}\chi_0 \\ \chi^-  \\ \chi^{--} \end{pmatrix}
    \qquad &\ \Rightarrow \qquad
    \chi_R t_R = ({\utilde{\bf 3}}, {\utilde{\bf 3}}, -1/3) \ .
\esp\ee
For the sake of the example, we focus on the first option where monotop
systems can be produced via the production and the decay of
a $\varphi_s$ resonance.
The presence of charged degrees of freedom in the $\chi$ multiplet
allows one to constrain this scenario
by other sources like, for instance, single bottom production
$\bar d_R \bar d_R \to \varphi_s \to b \chi^+$.
The mass splitting between the neutral and charged component of
the $\chi_d$ doublet being generated by electroweak loop-diagrams
(unless they mix to other fermions),
the decay of the charged $\chi^+$ field is driven
by the two competing channels $\chi^+ \to \bar b_L \bar d_R^i \bar d_R^j$
(mediated by $\varphi_s$) and $\chi^+ \to \chi_0 W^*$ (with a very
off-shell $W$-boson).
As a result, the $\chi^+$ particle is in general long-lived, which
is heavily constrained by current
LHC searches~\cite{Aad:2013pqd,Chatrchyan:2013oca}.
Furthermore, $\chi_d$ has the same quantum numbers as the
lepton doublets of the Standard Model so that these fields can mix,
which induces the proton decay modes $p \to \pi^+/K^+ \nu$ and
$p \to \pi^0/K^0 e^+$ similarly as
described above. Consequently, it turns out that
scenarios where the invisible $\chi$ fermion is one of the components
of a larger $SU(2)_L$ multiplet are unlikely to be realized.
Although we will ignore those non-minimal
scenarios, their complete analysis is however in order,
which goes beyond the scope of
this work focusing on monotop production only.

\subsubsection*{Spin-1 mediator}
We now turn to cases where monotop systems are produced from the decay of
a spin-$1$ resonance $X$. The related Lorentz vector fermionic bilinear
is given by $\bar{\psi}\gamma_\mu \psi$,
which can be written as $\bar{\psi}_L \gamma_\mu \psi_L + \bar{\psi}_R \gamma_\mu \psi_R$.
This implies that the two quarks have the same chirality. In order to build a scalar invariant, vector fields have
to couple to these spinors of the same chiralities. Using the same properties of the charge conjugation used in the notations for the scalar case, 
the possible couplings of the $X$-field to down-type quarks are then of the form
\beq
  \lambda_V^1\ X^\mu \bar{d}_L^C \gamma_\mu d_R +
   {\rm h.c.} \ ,
\eeq
where we denote the coupling strength by $\lambda_V$. In order for such
couplings to be $SU(2)_L$-invariant, the $X$-boson must belong
to a weak doublet with hypercharge $1/6$,
\beq
  X_\mu  = ({\utilde{\bf 3}}, {\utilde{\bf 2}}, 1/6) \equiv
   \left( \begin{array}{c} X_\mu^{2/3} \\ X_\mu^{-1/3} \end{array} \right)\ .
\eeq

Turning to the final state, we begin with the fact that
the $X$-field defined above
has the quantum numbers of a left-handed quark doublet. It
can consequently couple to a left-handed top quark and
a singlet field $\chi$,
\beq
  \lambda_V^2 X_\mu \bar{q}_L \gamma^\mu \chi + {\rm h.c.}
\eeq
Enforcing weak isospin gauge invariance implies thus, in addition to
the interaction relevant for the production of a monotop
state, the presence of the
interaction of a left-handed bottom quark to the second
component of the $X$-doublet. This however induces the fast decay of the
neutral $\chi$ fermion via an off-shell $X$-state,
\beq
  \chi \to b_L (X_\mu^{-1/3})^* \to b_L u_L d_R
   \quad(\text{or}\ \ \bar b_L \bar u_L \bar d_R \text{ if }
   \chi\text{ is a Majorana fermion})\ ,
\eeq
so that this model does not predict any monotop signal.
We therefore move on with a second option for linking the $\chi$-$t$ monotop
system to the $X$-boson by considering right-handed quarks.
A coupling to right-handed quarks can be obtained if the fermion $\chi$ belongs to an $SU(2)_L$ doublet with hypercharge $1/2$, $\chi_d = ({\utilde{\bf 1}}, {\utilde{\bf 2}}, 1/2)$,
\beq
  \lambda_V^3 X_\mu \bar{t}_R \gamma^\mu \chi_d + {\rm h.c.}
\eeq
This model however contains a charged fermion that can be produced in association with a top via the vector of charge $-1/3$ and is therefore likely to be constrained
by channels different from the monotop one.

More complicated non-minimal possibilities could also be considered but will
be ignored from this work for the same reason: they are
strongly constrained by other processes and their analysis
must account for these other channels, in addition to the monotop signature.

\subsubsection*{Summary for the resonant channel}

From the analysis of gauge invariant effective Lagrangians performed in this
subsection, we have shown that the chirality of both the initial down-type
quarks and the final-state top quark are correlated with the quantum numbers
allowed for the bosonic mediator and the invisible fermion. In
this work, we focus on the minimal model in terms of field content
and interactions. In this case, the setup
that predicts monotop production at the LHC as its main signature (and that
is thus not constrained by any other observable) contains a scalar mediator
and a new fermion that are both gauge-singlet and couple to right-handed quarks.
The effective Lagrangian reads
\be\bsp
  \mathcal{L}_{\rm eff.} = {\cal L}_{\rm kin}(\varphi_s,\chi) +
    \lambda_s^{ij} \ \varphi_s\, \bar{d}_{R,i}^C d_{R,j}
    + y_s\ \varphi_s^\dagger \, \bar{\chi} t_R  + {\rm h.c.} \  ,
\esp \label{eq:effres}\ee
where ${\cal L}_{\rm kin}$ contains gauge interaction, kinetic and mass terms
for the new fields, and the other terms focus on their interactions with
the Standard Model quarks.
This can be compared to the notation of Ref.~\cite{Andrea:2011ws}
where the Lagrangian describing the same
scenario is written as
\be\bsp
\mathcal{L}_{\rm res} &\ = {\cal L}_{\rm kin}(\varphi_s,\chi)
    + \Big( \varphi\, \bar{d}^C_i \Big[ (a_{SR}^q)^{ij} + (b_{SR}^q)^{ij}\gamma^5\Big] d_j
    + \varphi\, \bar{t} \Big[ a_{SR}^{1/2} +  b^{1/2}_{SR} \gamma^5 \Big] \chi + {\rm h.c.} \Big) \ ,
\esp\label{eq:lres}\ee
flavour indices being noted by $i,j$ and colour indices being omitted for clarity.
The two Lagrangians are related by
\beq
a_{SR}^q = b_{SR}^q = \lambda_s/2\, , \qquad a_{SR}^{1/2} = b^{1/2}_{SR} = y_s^*/2\,.
\eeq
As already mentioned, the couplings of the scalar to the down quarks are antisymmetric under the exchange of the flavour
indices. Consequently, parton density effects enhance the
production mode $d s \to \varphi^\ast$ at hadron colliders (with the relevant
coupling strengths being non-vanishing), as already
pointed out in previous works~\cite{Andrea:2011ws,Fuks:2012im,Agram:2013wda}.

\subsection{Non-resonant monotop production}
\label{sec:lnonres}
In the second class of scenarios implying the production of monotop states, the
top quark is produced in association with an invisible bosonic field that couples
in a flavour-changing way to top and light up-type (up or charm) quarks.
The bosonic state is in general not stable since it
couples to quarks. A missing energy signature is therefore enforced by requiring
these fields either to be long-lived so that they decay outside of the detector,
or to decay predominantly into a pair of additional neutral stable particles. In particular,
the latter possibility has been proposed in the framework of flavourful dark
matter models~\cite{Kamenik:2011nb}, where the extra boson ($\phi$ or $V_\mu$) is
a mediator of the interactions of the dark matter candidate with the Standard
Model particles.

The main issue with this class of models is to ensure that the new boson leads to
a missing energy signature in a detector. In this work, we address it by
assuming that the $\phi/V$ field dominantly decays into a pair of dark matter
candidate particles.\footnote{One could
also consider that neither the boson nor its decay products are stable, but instead
long-lived. Although this is a viable assumption, this implies further
complications in the building of the model. We therefore stick with the minimal
case.} In this case, extra constraints arise
from the requirement that the particle the boson decays into is a good candidate
for dark matter, or that at least it does not overpopulate the
Universe.

\subsubsection*{Spin-0 mediator}

As already stated in Section~\ref{sec:lres}, the interactions of a scalar field
to quarks involve both the right-handed and left-handed components of the
fermions. 
Consequently, the scalar $\phi$ field must transform as a doublet of $SU(2)_L$
with an hypercharge quantum number of $1/2$,
\beq
  \phi =({\utilde{\bf 1}}, {\utilde{\bf 2}},1/2)
     \equiv \left( \begin{array}{c} \phi^+\\ \phi^0\end{array} \right) \ .
\eeq
The coupling in the effective Lagrangian can be written as
\beq
  y^{ij}\ \phi \, \bar{q}_{L,i} u_{R,j} + {\rm h.c.} \label{eq:phinr}
\eeq
where $i,j$ span over the quark flavours.
The new scalar has the same quantum numbers as the Brout-Englert-Higgs field, thus one can also write a coupling to the right-handed down-type quarks and mix $\phi$ with the Higgs with the potential harm of generating a
non-vanishing vacuum expectation value. The presence alone of couplings to both
up-type and down-type quarks already generates dangerous flavour-changing effects.
Nevertheless, we assume, in a first step, that the only extra coupling
with respect to the Standard Model is the one of
Eq.~\eqref{eq:phinr} and will show, in the following,
that it is already hard to construct a phenomenologically viable
model. Gauge invariance implies the
presence of interactions between the charged component field $\phi^+$ and
quarks, so that the $\phi^+$ field always promptly decays into
two-body final states, $\phi^+ \to u\bar{b}$ or $t\bar{d}$. Analogously, the
neutral component $\phi^0$ could also decay into an associated particle pair
comprised of a top and an up quark, $\phi^0 \to u\bar t+ t \bar u$, as well as into
a three-body final state via the exchange of a virtual charged scalar
field\footnote{Due to reasons already stated in Section~\ref{sec:lres},
the $\phi^+$ and $\phi^0$ states are assumed to have similar
masses.}, $\phi^0 \to W^- [\phi^+]^* + W^+ [\phi^-]^*
\to W^- \bar{b} u + W^+ b \bar{u}$ or $W^- \bar{d} t + W^+ d \bar{t}$.
All these decay channels
are however assumed to be negligible when compared to
a decay into a pair of dark matter particles. In this case, no minimal coupling
to a single stable state is achievable  since $\phi$ is a doublet of
$SU(2)_L$, and one
must design an interaction of the $\phi$ state to two extra fields whose
combination forms a doublet of $SU(2)_L$. If we restrict ourselves to $\phi^0$-decays into
fermionic particles, the most minimal option is given by the Lagrangian
\beq
  {\cal L}_{\phi-{\rm decay}} = y_{\chi}\, \phi \bar{\chi_d} \chi_s + {\rm h.c.} \ ,
\eeq
where $\chi_s$ is an electroweak singlet and $\chi_d$ a weak doublet with
an hypercharge of $1/2$.
This term induces decays of both components of $\phi$
\beq
  \phi_0 \to \chi_s \chi_d^0
  \qquad\text{and}\qquad
  \phi^+ \to \chi_d^+ \chi_s \to [W^+]^* \chi_d^0 \chi_s\ ,
\eeq
the charged component $\chi_d^+$ being taken heavier than, but close in mass to,
the neutral component $\chi_d^0$ so that both neutral fields $\chi_s$ and
$\chi_d$ can be seen as viable dark matter candidates.

As a consequence of this non-minimal dark sector of the model, monotop
production via flavour-changing interactions of up-type quarks with a new
invisible scalar field will always be accompanied by an extra single top
production mode
\beq
  p p \to t \phi^- \to t \chi_d^0 \chi_s^0 [W^-]^*\ .
\label{eq:LFCSnewproc}\eeq
The nature and magnitude of the associated effects are very
benchmark dependent. For instance, a small mass splitting between the
component fields of $\chi$ leads to very soft $W$-boson decay products, so
that the process of Eq.~\eqref{eq:LFCSnewproc} would imply new contributions
to monotop production. On the other hand, in the case of larger mass splittings,
related new physics scenarios feature an LHC signature comprised of a single top
quark and an isolated lepton.

Nevertheless, we choose to keep the focus on minimal models, and therefore ignore, in the rest of this work,
scenarios where monotop states are produced from flavour-changing
interactions of up-type quarks with a 
scalar particle mediating dark matter couplings to the Standard Model.

\subsubsection*{Spin-1 mediator}
When the mediator is a vector boson $V$, one can design very simple models since
it can be singlet under the electroweak group. In this setup,
the associated couplings involve either
right-handed or left-handed quarks and take the form
\beq
  \Big(a_R^{ij}\ V_\mu \bar{u}_{R,i} \gamma^\mu u_{R,j} +
  a_L^{ij}\ V_\mu (\bar{u}_{L,i} \gamma^\mu u_{L,j} + \bar{d}_{L,i} \gamma^\mu d_{L,j}) +
  {\rm h.c.} \Big) \ ,
\label{eq:spin1coup} \eeq
where the $a_{L,R}$ parameters denote the strengths of the interactions
of the $V$-field with the quarks. As in the rest of this section, we
restrict ourselves to interactions focusing on
the monotop hadroproduction modes, so that only couplings involving
the third generation are assumed to be present. Moreover, our analysis
will focus on monotop production modes enhanced by parton densities. As a
consequence, interactions to second generation quarks are ignored.
Furthermore, being a singlet, $V$ can mix via a kinetic term with the hypercharge gauge boson in the Standard
Model: such mixing will in turn generate couplings of $V$ with all the quarks and leptons proportionally to their hypercharge.
As a result, bounds on the new particle $V$
can be derived from many new production and decay processes, like for intance,
its Drell-Yan production followed by a dilepton decay.
For this reason, we ignore possible kinetic mixing of the $V$ field
in the following.

The Lagrangian terms of Eq.~\eqref{eq:spin1coup} open various decay channels
for the $V$-field. Recalling that only couplings involving
the first and third generation quarks have been retained,
non-vanishing left-handed couplings allow the mediator to promptly
decay into jets initiated by down-type quarks,
$V \to b \bar{d} + d \bar{b}$.
Next, the importance of the decays into top and
up quarks (this time both in the context of left-handed and right-handed couplings) depends
on the mass hierarchy between the mediator and the top quark, the tree-level decay
$V \to t \bar{u} + u \bar{t}$ being only allowed when $m_V > m_t$. Furthermore, when
$m_V < m_t$, a triangle loop-diagram involving a $W$-boson could also
contribute to the
decay of the $V$-field into a pair of jets, $V \to d_i \bar{d}_j$. Finally, when
$m_W < m_V < m_t$, the three-body decay channel $V \to b W^+ \bar{u} + \bar{b} W^- u$ is open,
mediated by a virtual top quark.
A monotop signal is thus expected only when the $V$-field is invisible and
dominantly decays into a pair of dark matter particles.
Since $V$ is an electroweak singlet, the associated couplings can be written,
in the case of fermionic dark matter, as
\beq
  {\cal L}_{V-{\rm decay}} =
    V_\mu \Big(g_{R \chi}\; \bar{\chi}_R \gamma^\mu \chi_R +
       g_{L\chi} \; \bar{\chi}_L \gamma^\mu \chi_L \Big)\ ,
\eeq
where $\chi$ is a Dirac fermion, singlet under the Standard Model gauge
symmetries. The consistency of the model, \textit{i.e.}, the requirement that
$V$ always mainly decays into a pair of $\chi$-fields and not into one
of the above-mentioned visible decay modes, implies constraints on the
Lagrangian parameters. They will be studied in details in the next section, together with
other requirements that can be applied to viable
non-resonant monotop scenarios.

For completeness, one can also couple the $V$-boson to
left-handed quarks in (non-minimal) scenarios where it lies
in a triplet of $SU(2)_L$, $V_t = ({\utilde{\bf 1}}, {\utilde{\bf 3}}, 0)$.
In this case, it is however difficult to build couplings to a
minimal dark matter sector, the simplest case being the one of a
fermionic doublet of $SU(2)_L$. This also predicts the existence
of a charged component that can be produced, at the LHC, in association
with a single top quark or that can give rise to
a monobottom signature (for small mass gaps among
the vector degrees of freedom).
Following the minimality principle, we will not consider this case any further.

\subsubsection*{Summary for the non-resonant channel}

Summarising all the considerations above, the minimal gauge-invariant Lagrangian
yielding monotop production in the flavour-changing mode is given by
\be\bsp
  \mathcal{L}_{\rm non-res} &\ = {\cal L}_{\rm kin}(V,\chi)
   +
    V_\mu \Big(g_{R \chi}\; \bar{\chi}_R \gamma^\mu \chi_R +
       g_{L\chi}\; \bar{\chi}_L \gamma^\mu \chi_L \Big)\\ &\ +
  \Big(a_R^{ij}\ V_\mu \bar{u}_{R,i} \gamma^\mu u_{R,j} +
  a_L^{ij}\ V_\mu (\bar{u}_{L,i} \gamma^\mu u_{L,j} + \bar{d}_{L,i} \gamma^\mu d_{L,j}) +
  {\rm h.c.} \Big)\ ,
\esp\label{eq:lnonresfinal}\ee
where the first term contains kinetic, mass and gauge interaction terms
for the $V$ and $\chi$ fields. In the notations of
Ref.~\cite{Andrea:2011ws,Agram:2013wda}, the second line of the Lagrangian
reads
\be\bsp
  \mathcal{L}_{\rm non-res} = &\ {\cal L}_{\rm kin} + \Big( V_\mu\, \bar{u}_i \Big[ (a^1_{FC})^{ij} \gamma^\mu + (b^1_{FC})^{ij} \gamma^\mu \gamma^5 \Big] u_j  + {\rm h.c.} \Big) \ ,
\esp\label{eq:lnonres}\ee
so that the two parameterizations are related by
\be
  a_{FC}^1 = \frac{a_R + a_L}{2}\,, \quad  b_{FC}^1 = \frac{a_R - a_L}{2} \ .
\ee
The two parameter bases
are therefore equivalent, although gauge invariance imposes
that non-vanishing left-handed couplings relevant for monotop
production are accompanied by
interactions with left-handed down-type quarks too.
Since such interactions also enable the production of mono($b$-)jet final
states,
a full analysis of this scenario should account for monojet search results.
In the following, we will mainly focus on the case $a_L = 0$ unless specified.
Note also that in general $\chi$ may be a Majorana fermion, however this is a less likely situation as, in order to couple to $V$, the dark matter candidate is expected to carry a $U(1)$ charge.
In the following, we limit ourselves to the Dirac case, but the results in the Majorana case are qualitatively similar.


\section{Monotop phenomenology specific to non-resonant models}
\label{sec:nonrespheno}
Some features of the resonant models mediated by a scalar, like the lifetime of the invisible fermion produced in association 
with the top quark, have been studied in details in
Ref.~\cite{Wang:2011uxa}. In the following, we therefore focus on various
features of non-resonant spin-1 models by studying the effective lifetime of the invisible
vector, associated single top signals, and the dark matter relic density.
We separately consider two regions of the parameter space which have very different
phenomenology: the case where the mediator is lighter than the top quark
(its mass $m_V$ being smaller than the top mass $m_t$) and the case where it
is heavier, with $m_V > m_t$.
On the basis of the minimality argument employed in the
previous section, we also restrict ourselves to the case where
monotop production is the only expected new physics signature of the
model. We therefore set $a_L = 0$ in the effective Lagrangian of
Eq.~\eqref{eq:lnonresfinal}. As stated above, non-vanishing
$a_L$ values imply mono($b$-)jets production, and the associated
constraints may be predominant. The corresponding
detailed study is postponed to future work.

\subsection{Mediators heavier than the top quark}

We first start with the scenario of heavy mediators: the mediator $V$ is not long lived as it can
always decay into a top quark.
Including in the model a $V$-decay channel into an invisible state to be
considered as a dark matter candidate
is thus always necessary. 
Focusing on the minimal case, we study below the interesting interplays between the
requirement that the invisible channel dominates and bounds originating
from the relic density of the dark matter candidate.

\subsubsection{Tree-level decays of the mediator}
\label{sec:LOdecays}
When the $V$-boson is heavier than the top quark, it can decay into
either a pair of down-type quarks, an
associated pair comprised of a top quark and a lighter quark or a pair of dark
matter particles, as already discussed in Section~\ref{sec:lnonres}. Since the
first two decay modes are driven by the same interaction vertices allowing one
for monotop production, we need to make sure that the invisible decay channel
always dominates. 
As we focus on the couplings to up and top quarks, we set all the couplings $a_{R,L}^{ij} = 0$ except for $a_{R,L}^{13}$ and $a_{R,L}^{31}$.
The relevant partial widths are given by\footnote{The results have been checked using the decay module
of {\sc FeynRules}~\cite{Alwall:2014bza}.}
\be\bsp
 \Gamma (V \!\to\! b \bar{d} \!+\! \bar{b} d) = &\ \frac{m_V}{4 \pi} |a_L^{31} + a_L^{13,\ast}|^2 \ , \\
 \Gamma (V \!\to\! t \bar{u} \!+\! \bar{t} u) = &\
    \frac{m_V}{4 \pi} |a_R^{31} + a_R^{13,\ast}|^2 \left(1-\frac{m_t^2}{m_V^2} \right) \left( 1-\frac{m_t^2}{2 m_V^2} - \frac{m_t^4}{2 m_V^4}  \right)\ , \\
 \Gamma (V \!\to\! \chi \chi) = &\
    \frac{m_V}{24 \pi} \sqrt{1\!-\! 4 \frac{m_\chi^2}{m_V^2}}\ \Bigg[ 
    \Big(|g_{L\chi}|^2\!+\!|g_{R\chi}|^2\Big) \Big(1\!-\!\frac{m_\chi^2}{m_V^2}
       \Big) +
    \frac{6 m_\chi^2}{m_V^2} \Re \{g_{L\chi} g_{R\chi}^*\} \Bigg]\ ,
\esp\label{eq:widths}\ee
where we neglect all quark masses but the top mass. In addition,
we denote by $m_\chi$ the mass of the dark matter candidate.
In the minimal scenario where only right-handed couplings are present, $a_L = 0$, the decay to light quarks vanishes and we are left with two decay channels above the top threshold.
For future convenience, we define $a_R = a_R^{31} + a_R^{31,\ast}$.

\begin{figure}
\centering
\includegraphics[width=.47\columnwidth]{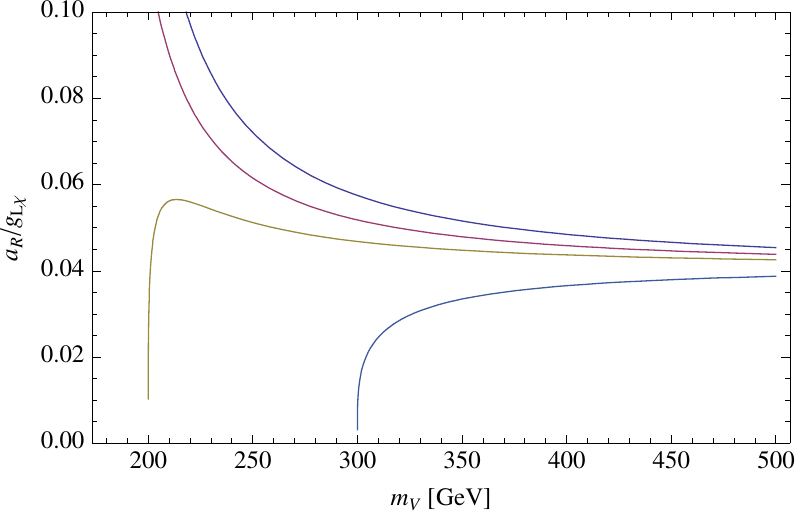}
\includegraphics[width=.47\columnwidth]{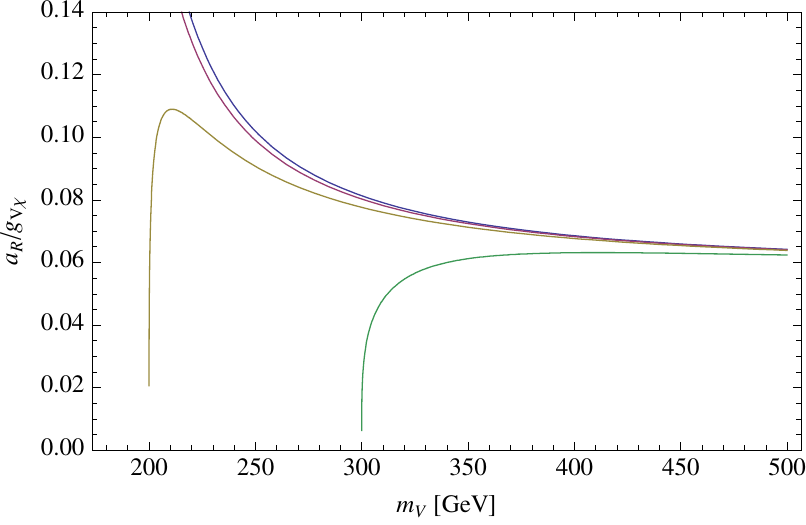}
\caption{Maximum value of $a_R$ necessary to enforce the mediator $V$ to decay
  invisibly in 99\% of the cases. We focus on scenarios where the couplings
  of the mediator to dark matter are chiral with $g_{R\chi}=0$ (or
  $g_{L\chi}=0$) in the left panel, and vector with $g_{L\chi} = g_{R\chi} =
  g_{V\chi}$ in the right panel. The four curves correspond to $m_\chi = 5$, $75$, $100$ and $150$~GeV from
  the lower to the upper ones in each figure.} \label{fig:V2body}
\end{figure}

In this set-up, we study typical
constraints that can be imposed on ratios of the $g_{L\chi}$, $g_{R\chi}$ and $a_R$
parameters when they are all assumed to be real quantities. Since ratios of
branching ratios are equivalent to ratios of partial widths, we use this latter
quantity and show, in Figure~\ref{fig:V2body}, the maximum value of the
$a_R$ coupling
strength in units of the $\chi V$ coupling that ensures the $V$-field to decay
invisibly in at least 99\% of the cases. In the left panel of the figure, we
consider scenarios where $g_{R\chi}$ vanishes (the same result holds for vanishing
$g_{L\chi}$), while
in the right panel of the figure, we assume vector-like couplings, $g_{L\chi} = g_{R\chi} = g_{V\chi}$. In general,
the coupling to the top quark $a_R$ (that is responsible for the monotop signal) has to be quite small compared to the coupling to the dark matter candidate in order for the mediator $V$ to be invisible, unless the mass of the mediator $V$ is close to the top mass.
On the contrary, if the mass of $V$ is close to the $\chi \chi$ threshold, the invisible decay modes are suppressed.
This study shows that it is not straightforward to have $V$ to decay invisibly, and this constraint may play an important role in the interpretation of the signal, especially when associated with the study of the properties of $\chi$ as a dark matter candidate. We study more in detail this question in the next subsection.

Similar conclusions would hold in less minimal models, like the one with a left-handed coupling $a_L$ where the decay to down-type quarks is open and dominant also below the top threshold.

\subsubsection{Dark matter constraints} \label{sec:DMheavy}

We have seen that, in order to avoid visible decays of the mediator $V$, it has to be
coupled to a stable particle $\chi$ and the decay $V \to \chi \chi$ must always dominate.
If $\chi$ is stable, and if the model is minimal in the sense that $V$ is the only mediator of interactions between the dark sector and the Standard Model, then the only annihilation process that will determine the thermal relic abundance of $\chi$ is $\chi \chi \to V \to t \bar{u}$ and $\bar{t} u$.
Such process is proportional to the same coupling that gives rise to the monotop signature at the LHC, and also to the coupling of $V$ to dark matter.
By studying the relic abundance of $\chi$ one can therefore derive interesting constraints on the couplings, especially when imposing that the relic abundance is smaller than the measured density of dark matter.
Those restrictions can in principle always be evaded by assuming that there are additional
mediators, or that $\chi$ is not a stable particle but rather a long-lived one that decays on cosmological time scales.
In the rest of the section, we nevertheless focus on the minimal case of $\chi$ being the only
dark matter candidate.

\begin{figure}[tb]
\begin{center}
\includegraphics[width=7cm]{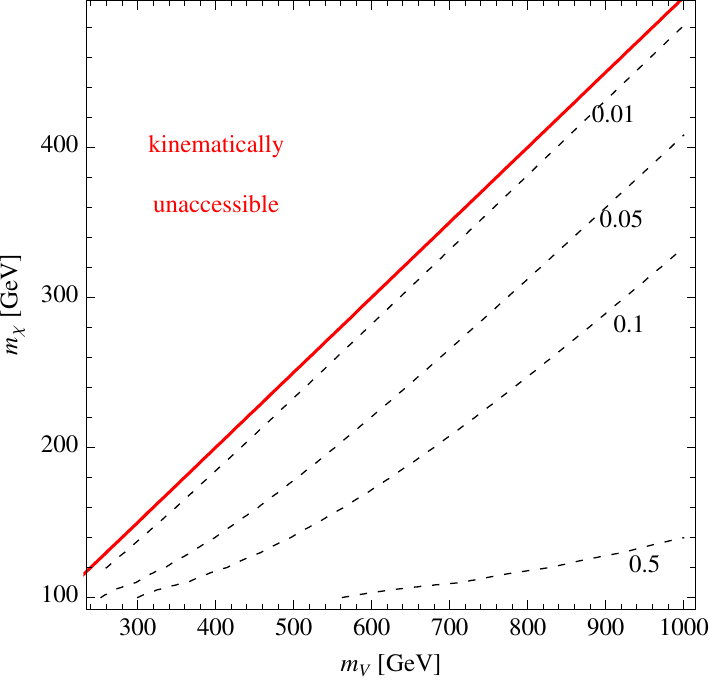}
\end{center}
\caption{Lower bound on $g_{V\chi} \times a_R$ from the dark matter relic abundance as a function of $m_V$ and $m_\chi$.} \label{fig:gVaR}
\end{figure}

As the relic abundance decreases with increasing annihilation cross sections, one can calculate a lower bound on the product of $a_R$
with the couplings of $V$ to the dark matter by requiring that the relic abundance is equal or smaller than the measured one. Values of the couplings below the bound would be excluded as the stable particle would overpopulate the Universe. The bound has been computed by implementing the model described by the
Lagrangian of Eq.~\eqref{eq:lnonresfinal} in {\sc CalcHep}~\cite{Belyaev:2012qa}.
For the calculation of the relic abundance, we used the usual approximate formulas deriving from an analytic solution of the Boltzmann equation (see Ref.~\cite{Kolb:1990vq} for more details):
\beq
\Omega_{DM} h^2 = \frac{1.04 \cdot 10^9}{M_{Pl}} \frac{x_F}{\sqrt{g_\ast}} \frac{1}{\langle \sigma v\rangle}
\eeq
where $x_F = m_{\chi}/T_F$ and the freeze-out temperature is $T_F \sim 25$~GeV, $g_\ast = 92$  is the number of relativistic degrees of freedom at freeze-out, and all dimensionful quantities are in GeV.
We consider, for concreteness, a vectorial model with $g_{L\chi} = g_{R\chi} = g_{V\chi}$.
The results of the calculation are shown in Figure~\ref{fig:gVaR}, where we present
the lower bound on $a_R \times g_{V\chi}$ as a function of the mediator mass $m_V$ and the dark matter mass $m_\chi$.
We restrict ourselves to values of the $\chi$ mass above the top threshold, $2 m_\chi > m_t$, so that a two-body process is kinematically allowed.
Below the top threshold, the dark matter candidate can only annihilate into three-body final
states or via loop-induced processes, so that the annihilation cross section is too small and the $\chi$ particle overpopulates the Universe.
The figure shows that the product of couplings is bound to be larger than about $0.1$, with the lower bound increasing towards the top threshold as the phase space closes down, and becomes smaller towards the $V$ threshold $2 m_\chi = m_V$ where the resonant $V$ exchange enhances the annihilation. We recall that the $V$-boson mass must be at least twice as large as the
dark matter candidate mass to allow invisible decays for $V$. The corresponding regions
of the parameter space are tagged as kinematically inaccessible.

\begin{figure}[tb]
\begin{center}
\includegraphics[width=7cm]{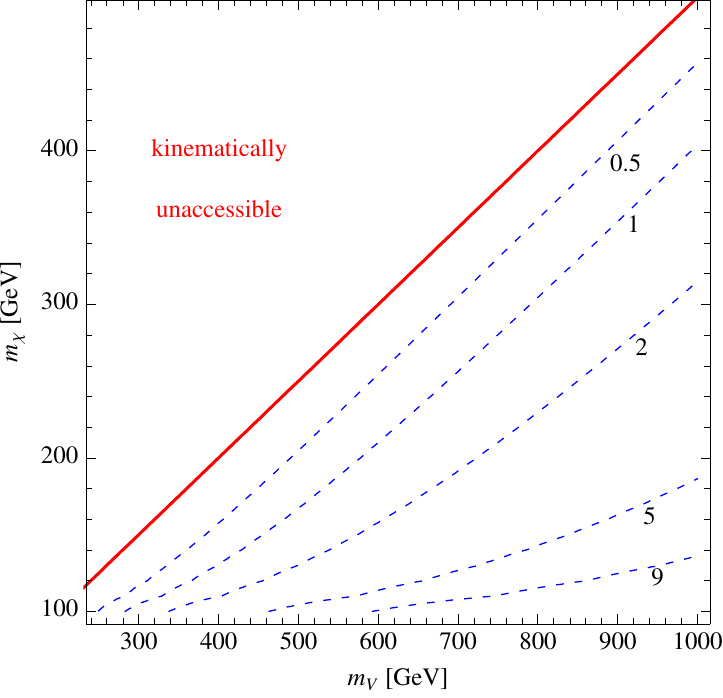}
\end{center}
\caption{Lower bound on $g_{V\chi}$ obtained combining the dark matter relic abundance
  constraints with the requirement that the mediator $V$ decays invisibly in 99\% of the cases.}
\label{fig:gV}
\end{figure}

This result, very interesting \textit{per se}, can be combined with other constraints to better determine
the viable regions of the parameter space of the model.
The requirement that the invisible $V$-decay dominates has allowed us, in
Section~\ref{sec:LOdecays}, to calculate a lower bound on the ratio $g_{V\chi}/a_R$ which
depends on the mediator and dark matter masses (see Figure~\ref{fig:V2body}).
Multiplying it with the limits derived from the relic abundance predictions, we
extract a lower bound on $g_{V\chi}$ independently of the value of $a_R$: the results are
shown in Figure~\ref{fig:gV}.
The lower bound on $g_{V\chi}$ is found to grow with smaller values of the $\chi$ mass.
Moreover, near the top threshold, it reaches values well above unity, tending hence to the
non-perturbative regime.

Under the assumption that $\chi$ is the only dark matter candidate of the theory, we can further
restrict our analysis to parameter space regions where the values of the couplings are such that
the bound from the dark matter abundance is saturated. We first reinterpret,
as a function of the masses, the limits calculated in the CMS monotop search~\cite{CMS:2014hba}
by accounting for an invisible branching ratio of the mediator that may not be 100\%.
Next, we correlate these to the dark matter results:
for increasing values of $a_R$, the coupling $g_{V\chi}$ has to be smaller to
satisfy the dark matter constraints. This indicates that an enhancement of the $tV$ production
rate (by increasing $a_R$) is accompanied by a reduction of the invisible branching
ratio of $V$, which possibly reduces the production cross section of monotop
systems.

\begin{figure}[tb]
\begin{center}
\includegraphics[width=7cm]{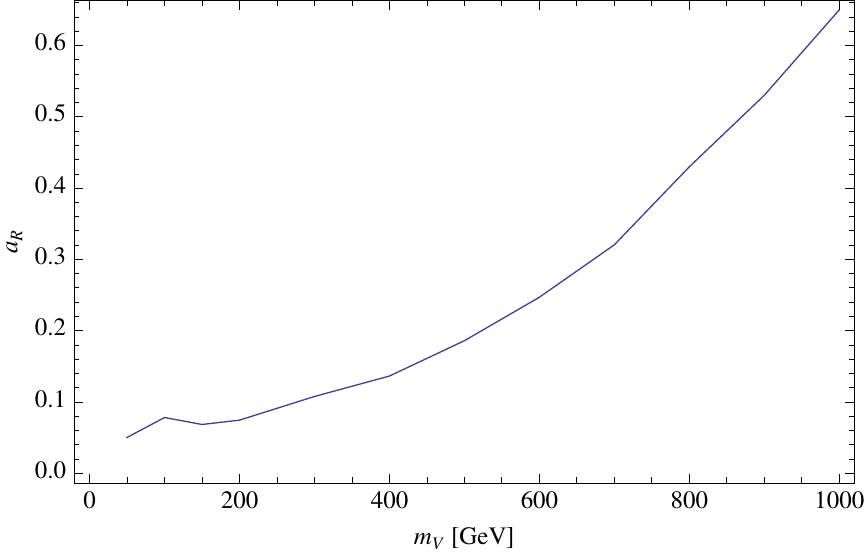}
\includegraphics[width=7cm]{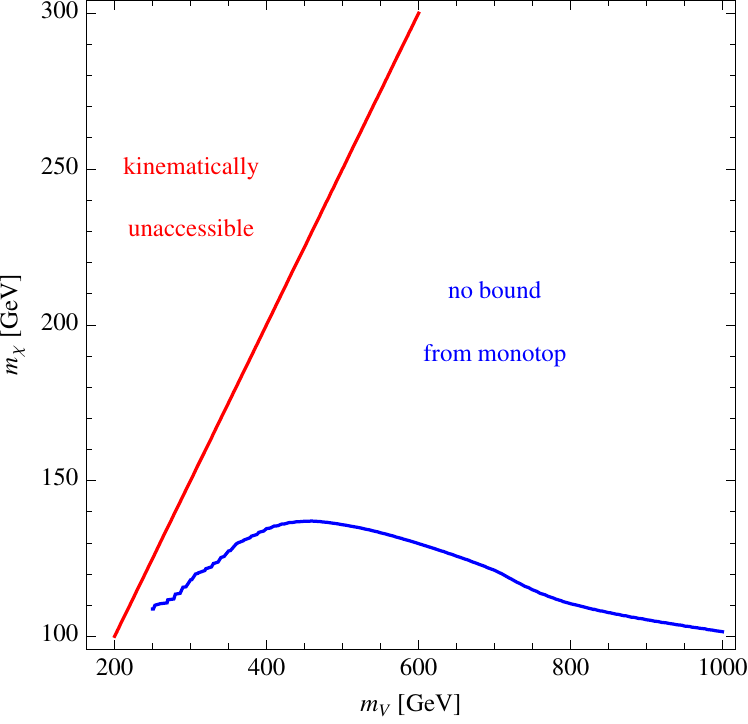}
\end{center}
\caption{Reinterpretation of the CMS monotop limits of Ref.~\cite{CMS:2014hba}
  in terms of $a_R$ (left) for a 100\% invisible mediator. The results are then used to determine the viable regions of the
  parameter space when enforcing dark matter and LHC constraints as shown by
  Eq.~\eqref{eq:arineq} (right). The region above the blue line is found not to be bounded by
  current searches. Below the blue line, limits on $a_R$ deduced from the left panel of the
  figure are in order.} \label{fig:DM}
\end{figure}

A general bound on $a_R$ can be obtained using the relation
\beq
a_R^2 \times \frac{k^2/a_R^2 \tilde{\Gamma}_{\chi\chi}}{k^2/a_R^2 \tilde{\Gamma}_{\chi\chi} + a_R^2 \tilde{\Gamma}_{tu}} \leq a_{R-CMS}^2 \ ,
\eeq
where $\tilde{\Gamma}$ denote the partial widths into $\chi \chi$ and $t u$ final states given
by Eq.~\eqref{eq:widths} stripped by the coupling strengths, $a_{R-CMS}$ is the upper bound on
$a_R$ derived from the CMS analysis that assumes that $V$ decays are always invisible, and
$k$ is the lower bound on $g_{V\chi} \times a_R$ deduced from the dark matter relic abundance
in Figure~\ref{fig:gVaR}.
On the left panel of Figure~\ref{fig:DM}, we extract the bound on $a_{R-CMS}$ from
the CMS analysis of Ref.~\cite{CMS:2014hba}.
Inverting the above equation, bounds on $a_R$ for a $\chi$ particle saturating the dark
matter relic abundance can then be rewritten as
\be\bsp
  &\ a_R^2 \leq \frac{k^2 \tilde{\Gamma}_{\chi\chi}}{2 a_{R-CMS}^2 \tilde{\Gamma}_{tu}} \left( 1 - \sqrt{1-4 \frac{a_{R-CMS}^4 \tilde{\Gamma}_{tu}}{k^2 \tilde{\Gamma}_{\chi\chi}}} \right) \ , \\
 \quad\text{or}\quad
&\ a_R^2 \geq \frac{k^2 \tilde{\Gamma}_{\chi\chi}}{2 a_{R-CMS}^2 \tilde{\Gamma}_{tu}} \left( 1 + \sqrt{1-4 \frac{a_{R-CMS}^4 \tilde{\Gamma}_{tu}}{k^2 \tilde{\Gamma}_{\chi\chi}}} \right)\,.
\esp\label{eq:arineq}\ee
The result is shown on the right panel of Figure~\ref{fig:DM}.
Above the blue curve, the argument of the square root is negative and the inequalities of
Eq.~\eqref{eq:arineq} have no solution, therefore there is no bound that can be applied on $a_R$.
Below the blue line, near the top threshold, the dark matter constraint requires larger couplings and therefore larger monotop rates are allowed, thus a bound on $a_R$ can
be calculated. Naturally, larger portions of the parameter space are expected
to be covered with the upcoming run II of the LHC.

The region where the monotop signal is suppressed can have interesting additional features.
The boson $V$ may dominantly decay into top and lighter quarks, yielding at the same time
a signature comprised of same-sign top quark pairs ($t V \to t t \bar{u}$) and extra
contributions to top-antitop production ($t V \to t \bar{t} u$) that may be difficult
to observe due to the overwhelming $t\bar{t}$ Standard Model background. These extra channels
deserve a particular attention, in particular in upcoming data from LHC collisions
at $\sqrt{s}=13$~TeV.

\subsection{Mediators lighter than the top quark}
\label{sec:phenolightv}

When the spin-1 mediator $V$ is lighter than the top quark, its possible decay
modes into a top and a lighter quark are kinematically forbidden.
At tree-level, in the minimal scenario $a_L = 0$, $V$ can therefore only decay into a multibody final state
such as $V \to u \bar{b} W^-$ or $\bar{u} b W^+$, where the $W$-boson is
virtual when $m_V < m_W$ ($m_W$ denoting the $W$-boson mass)~\footnote{When $a_L \neq 0$, the decays into a pair of down quark will always dominate.}. In
this mass range, loop-induced decays must however be considered too. For instance,
a triangle loop-diagram with a $W$-boson exchange
generates couplings to down-type quarks, thus opening a dijet decay channel.
As the decay channels in this region are either kinematically or loop-suppressed, one may wonder 
whether $V$ may be long-lived without the need 
for an additional invisible decay channel.
Another interesting property of this mass region is that a new decay of the top quark
is allowed, $t \to u V$, and extra constraints on monotop scenarios could
therefore be extracted
from, \textit{e.g.}, top width measurements or the analysis of $t \bar{t}$ events when one
of the top quarks decays into a jet plus missing energy.

\subsubsection{Loop-induced and multibody decays of the mediator}
\label{sec:loopdecays}

Light mediators, below the top mass threshold, may decay dominantly
into two jets via loop-induced interactions.
The structure of the loop crucially depends on the chirality of the monotop couplings: for instance, in the case $a_L \neq 0$, the loop is divergent,
which signals that a tree-level coupling to down-type quarks is necessary for the theory to be consistently renormalizable.
This result confirms the necessity to consider an effective model fully invariant under the electroweak symmetry.

In the minimal case $a_L=0$, the loop contributions turn out to be finite.
Since weak interactions are left-handed, the
chiralities of the quarks involved in these diagrams must be flipped, which
implies that the loop-induced couplings are proportional to the product of the
up and top masses $m_u m_t$. Contrary to setups where monotops are produced from
left-handed interactions of the mediator with quarks, the loop-induced $Vd_Ld_L$
couplings are thus finite, in line with the fact that no associated
counterterm appears after renormalization. The interaction strength reads, in
the limit of small light quark masses,
\beq
g_{Vd^id^j}^{\rm 1-loop} (a_R)= \frac{\alpha a_R}{4 \pi s_W^2} \frac{m_u}{m_t} (V_{ud^i}^* V_{td^j} + V_{td^i}^* V_{ud^j}) \tilde{c}_0\ ,
\eeq
where $\alpha$ stands for the electromagnetic coupling constant, $s_W$ for the
sine of the weak mixing angle and $V_{ij}$ for the elements of the CKM matrix.
In addition, the loop factor
\beq
  \tilde{c}_0 = m_t^2\ C_0 (p_1, -(p_1+p_2); m_W, m_t, 0)
\eeq
depends on the Passarino-Veltman three-point function $C_0$ where $p_1$ and
$p_2$ are the momenta of the external down-type quarks.
We can therefore calculate the partial width associated with the
decay $V\to\bar{d}^id^j$ which reads, after summing over
all down-type quark flavours,
\beq
  \Gamma (V \to jj) = \frac{\alpha^2 a_R^2}{64 \pi^3 s_W^4} \frac{m_V m_u^2}{m_t^2} |\tilde{c}_0|^2\ .
\label{eq:loopwidth}\eeq
We observe that it exhibits both a loop-suppression and a
$(m_u/m_t)^2$ factor, so that it is expected to be numerically small.

\begin{figure}
  \centering
  \includegraphics[width=.47\columnwidth]{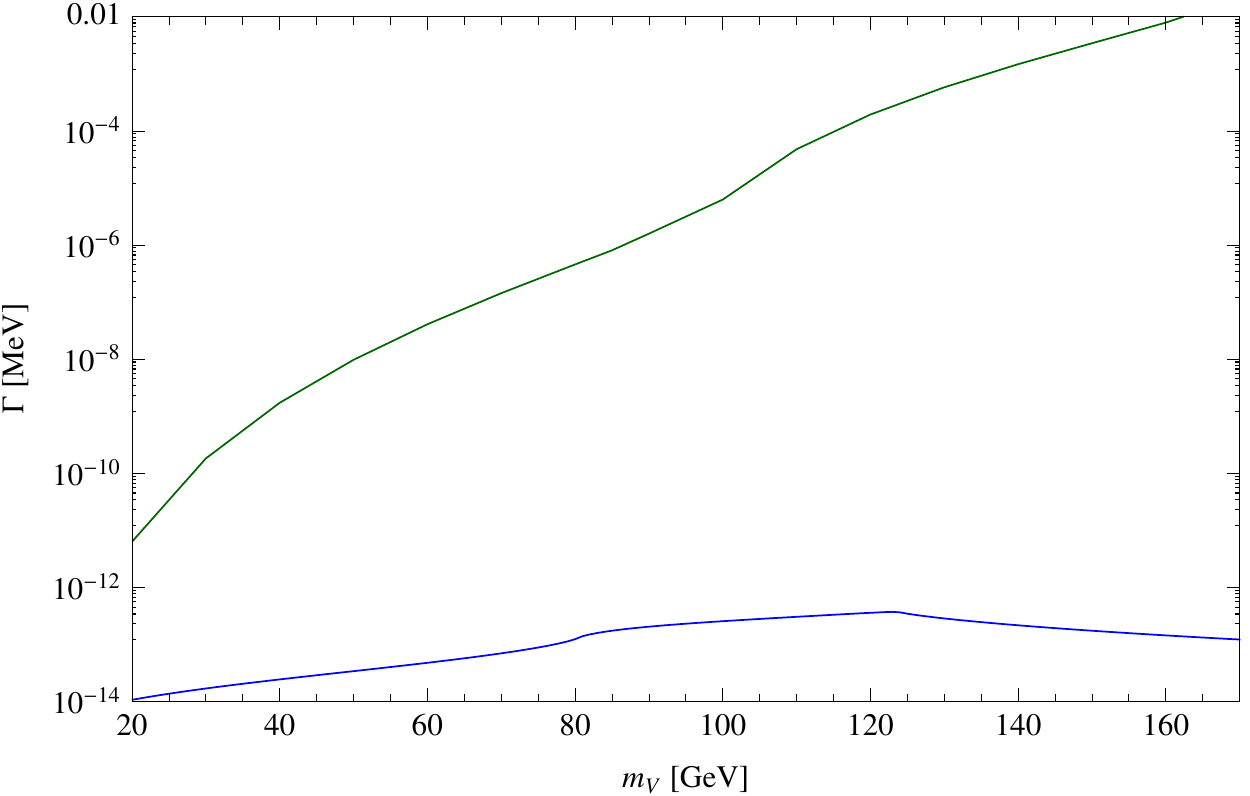}
  \includegraphics[width=.47\columnwidth]{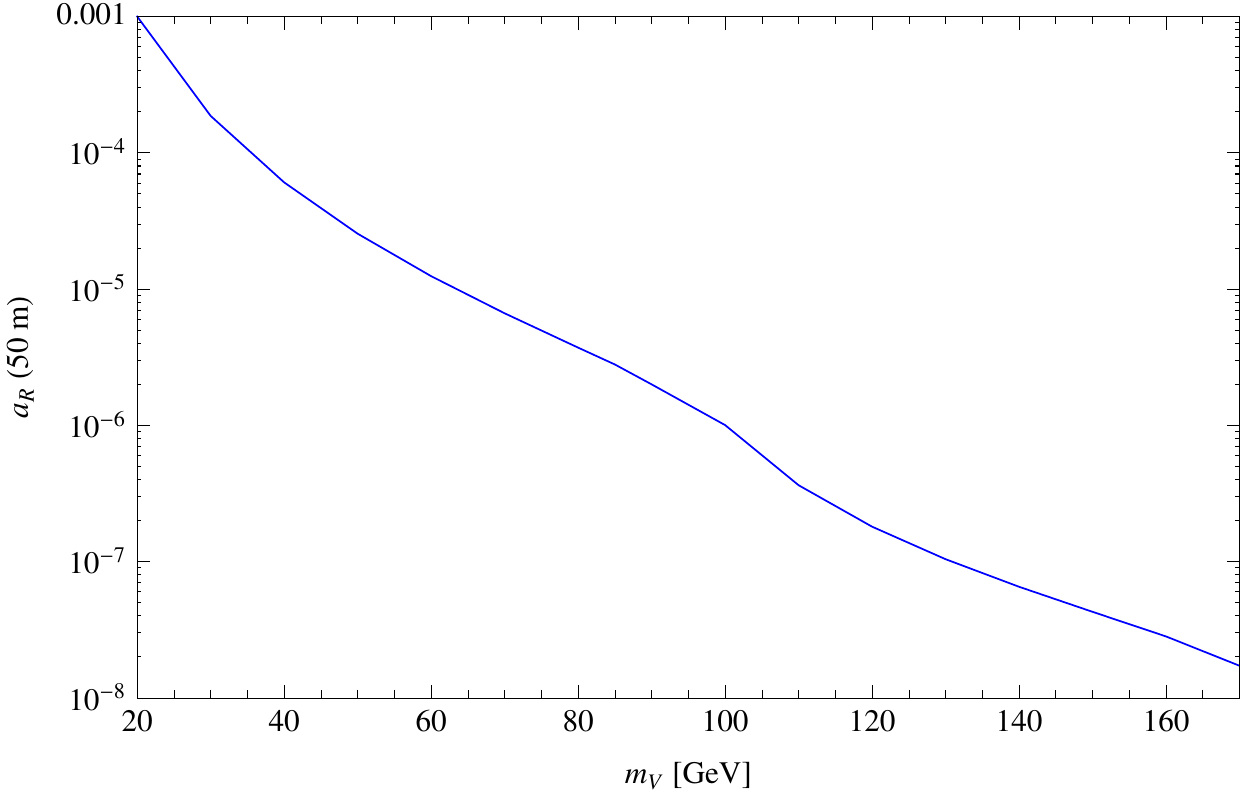}
  \caption{Partial decay width associated with the loop-induced (blue)
    and multibody (green) decay of the
    $V$-field into down-type quarks as a function of the $V$-boson mass, as
    given by Eq.~\eqref{eq:loopwidth} and returned by {\sc MadWidth} (left panel). We consider
    scenarios in which $a_R = 0.04$ and $m_V$ is kept smaller than the top mass.
    The results are translated, in the right panel, as a bound on $a_R$ that
    ensures that the $V$-boson has a decay length of at least 50 m.}
  \label{fig:V2bodyloop}
\end{figure}

In Figure~\ref{fig:V2bodyloop}, we show the partial width in Eq.~\eqref{eq:loopwidth} as a function of the mediator mass for $a_R = 0.04$ (left panel, blue curve).
We compare this result to preditions for three-body and four-body decays (left panel, green) as
calculated by {\sc MadWidth}~\cite{Alwall:2014bza}, which turn out to be dominant upon the entire
mass range.
On the right panel of the figure, the partial width is translated as an upper bound on the value of $a_R$ in order for $V$ to have a mean decay length of at least 50 metres so that it is long-lived enough to decay outside of typical hadron collider detectors.
The figure shows that the lifetime of $V$ would be long enough only for extremely small
values of the coupling $a_R$ that will challenge the possible
observation of a monotop signal at the LHC by reducing the associated production cross section.
The only way out is thus to extent the theoretical framework so that invisible decay channels
are enabled, as in the previous section.
It should also be mentioned that above the $W$-boson threshold, a tree-level three-body decay is kinematically open, which further shortens the decay length of $V$. Finally, in cases where the model features a coupling of the $V$-field to
top and charm quarks, the
partial width of Eq.~\eqref{eq:loopwidth} would exhibit an enhancement
proportional to $(m_c/m_u)^2$. 

In summary even for monotop scenarios in which the mediator cannot decay into a top quark,
its lifetime is generally too short and one needs to complete the model by adding a decay channel into an invisible state.
Although the class of minimal scenarios described in this section features a light extra
vector boson, the setup is compatible with current Tevatron and LHC bounds on
monotop production as the latter are always derived under the assumption of very
large coupling values of ${\cal O}(0.1)$~\cite{Aaltonen:2012ek,CMS:2014hba}.
They could however be constrained by other observations, as will be shown in
the next subsections.

\subsubsection{Single top constraints on monotop scenarios}

\begin{figure}
  \centering
  \includegraphics[width=.5\columnwidth]{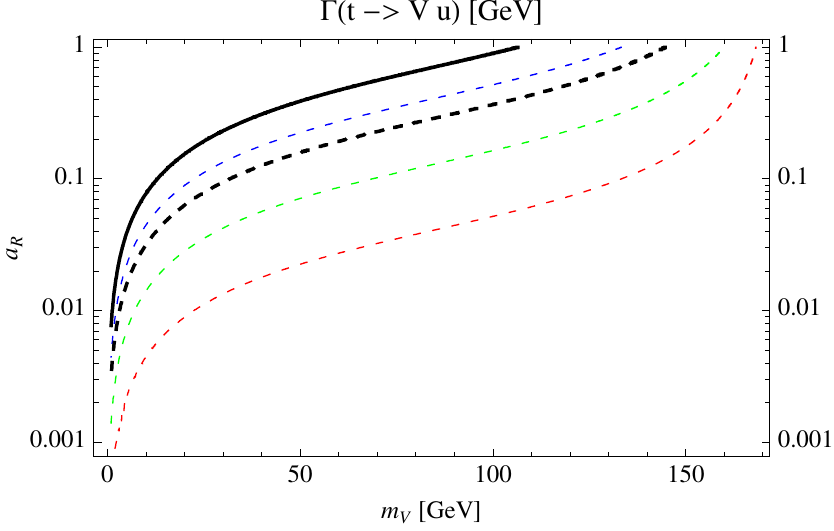}
  \caption{Partial width associated with the $t\to V u$ decay mode of the
    top quark as a function of the $V$-boson mass and the $a_R$ coupling. We show
    curves for a partial width of 3, 1, 0.5, 0.1 and 0.01~GeV. The solid black curve corresponds to the upper bound on $a_R$ from a partial width less than $3$~GeV, roughly corresponding to the direct measurement in Ref.~\cite{Aaltonen:2013kna}.} 
  \label{fig:GammaTop}
\end{figure}

Motivated by minimality principles, we have discussed, in the previous section,
appealing monotop scenarios in which the mediator $V$ is lighter than the top
quark. In this case, the
former couples to up and top quarks via right-handed couplings and one needs to add an invisible decay channel to potential dark matter particles $\chi$ to guarantee a monotop signature,
unless the coupling strength $a_R$ is very small. On different
grounds, these scenarios feature a new decay channel for the top quark,
$t \to u V$. This observation can be used to further restrict the viable regions
of the parameter space by imposing that new
physics contributions to the top width do not challenge the measured value of
$1.10<\Gamma_t < 4.05$~GeV~\cite{Aaltonen:2013kna}, from direct measurements at CDF.
A more precise measurement, which takes the value $\Gamma_t = 2.0 \pm 0.5$~GeV~\cite{Beringer:1900zz}, can also be obtained by fitting the single-top measurements.
However, the latter does not apply in our scenario where new physics contributions to single-top can arise. 
Assuming a good agreement between the Standard Model expectation and the top width measurement, therefore, the partial
width $\Gamma(t\to V u)$ can thus be enforced to be of at most 3~GeV. On
Figure~\ref{fig:GammaTop}, we present the dependence of this partial width on the
coupling $a_R$ and the mediator mass $m_V$. We observe that for couplings smaller
than 0.01, new physics effects in the top width are predicted to be very small,
except when the mediator is almost massless. This consequently disfavours such
setups in which the mediator is very light, even in cases with coupling strengths
of ${\cal O}(0.001)$.

Kinematically allowed $t\to Vu$ decays also imply that monotop events can be
issued from the production of a top-antitop pair when one of the top quarks decays
into a $V$-boson and a light quark,
\beq
  pp \to t \bar{t} \to t \bar{u} V \qquad\text{or}\qquad
  pp \to t \bar{t} \to \bar t u V \ .
\eeq
This process induces additional contributions to the production of a monotop system ($tV$ or
$\bar t V$) in association with an additional jet, a signature already
accounted for in the LHC monotop analysis of Ref.~\cite{CMS:2014hba}. 
How much this new channel will contribute to the monotop signal depends on the cuts employed  
in the experimental analysis. However, due to the large $t\bar{t}$ cross section,
these effects cannot be neglected.

Complementary constraints on this channel could be deduced from Standard Model single top
analyses whose signal regions could capture monotop events as above. For instance,
both CMS~\cite{Khachatryan:2014iya} and ATLAS~\cite{Aad:2012ux} have analyses dedicated to the measurement of the single top cross section in
the $t$-channel which contain
a region that could be populated by monotop events as
above.\footnote{Other single top analyses could be
considered. However, they in general use multivariate techniques that cannot be
employed in the reinterpretation framework pursued below.}
In the CMS analysis, events are selected by requiring one
single isolated electron or muon and exactly two jets, one of them being
$b$-tagged. The background is reduced by requiring an important amount of missing
energy and by imposing that the transverse mass computed after combining the
lepton transverse momentum with the missing transverse momentum is large.
A final selection is preformed by means of an advanced multivariate technique.
We have nevertheless to ignore this last step of the selection as the
amount of information provided in the experimental publication is not sufficient
for satisfactorily recasting it (see Ref.~\cite{Kraml:2012sg} for more
information on this aspect).

\begin{figure}
  \centering
  \includegraphics[width=.5\columnwidth]{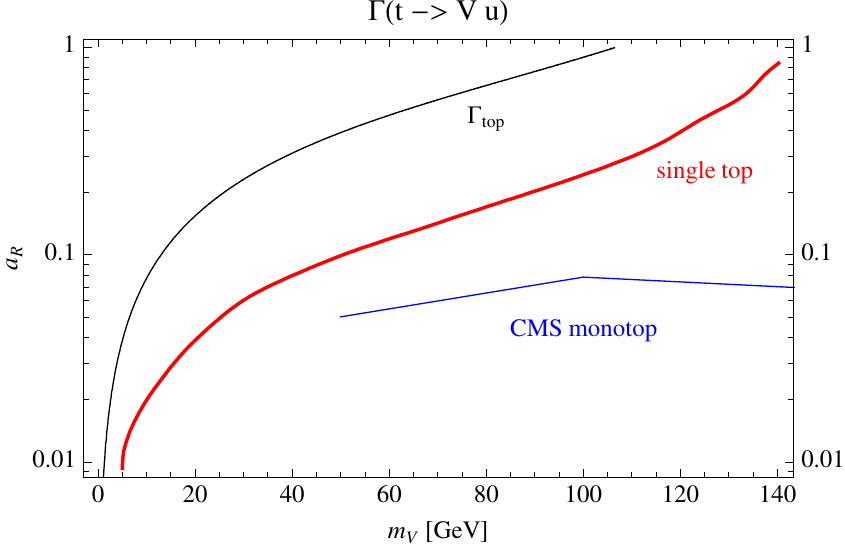}
  \caption{Parameter space region excluded (the area above the red curve)
    at the 95\% confidence level by the
    CMS single top analysis of Ref.~\cite{Khachatryan:2014iya} before the multivariate selection, compared to the exclusion from the top width~\cite{Aaltonen:2013kna} (black line), and the CMS monotop search~\cite{CMS:2014hba} (blue line).}
  \label{fig:cms12038}
\end{figure}

We simulate our new physics signal by using the monotop model~\cite{Andrea:2011ws}
implemented in the {\sc FeynRules} package~\cite{Christensen:2008py,Alloul:2013bka},
tuning the model parameters to the setup of Eq.~\eqref{eq:lnonresfinal}, so that
we can export the model to a UFO library~\cite{Degrande:2011ua} that is then
linked to
{\sc MadGraph5\_aMC@NLO}~\cite{Alwall:2014hca}. The generated parton-level events
have subsequently been processed by {\sc Pythia}~\cite{Sjostrand:2006za} for
parton showering and hadronization and by {\sc Delphes}~\cite{deFavereau:2013fsa}
for detector simulation, making use of the recent `MA5Tune'~\cite{Dumont:2014tja}
of the CMS detector description of {\sc Delphes}. The CMS analysis of
Ref.~\cite{Khachatryan:2014iya} has finally been implemented in the
{\sc MadAnalysis5} framework~\cite{Conte:2012fm,Conte:2014zja}, which has allowed
us to derive exclusion bounds at the 95\% confidence level in the $(m_V,a_R)$
plane, as shown on Figure~\ref{fig:cms12038}. 
The figure also shows the constraint from the top width, and from the dedicated CMs monotop search~\cite{CMS:2014hba}.
The monotop search is currently more sensitive.
However, the bound from the single top is a rough estimate, and the bound may be much stronger once the full analysis, including
the multivariate selection, is taken into account.
Nevertheless, our result shows that the constraints from single top searches can play an important role in constraining monotop scenarios.
The only region in parameter space where the monotop search will always be dominant is the region where the vector is close in mass to the top, because the single top channel will be suppressed by phase space while the monotop signal is not.

\subsubsection{Dark matter constraints}

We have argued that, even for mediator masses below the top threshold, an invisible decay channel is typically needed in order for the monotop signature to be present.
The simplest way out is to couple $V$ to a fermionic stable dark matter candidate $\chi$.
However, in a minimal scenario where $V$ is the only mediator for the interactions of the dark matter candidate, one needs to ask whether the relic abundance of $\chi$ is enough to fulfill the bounds from observations.
Below the top threshold, the main annihilation process $\chi \chi \to V \to t \bar{u}$ and $\bar{t} u$ is kinematically forbidden,
so that the annihilation of dark matter particles can only proceed to a
three-body or four-body final state (via a virtual top quark), or via loop-diagrams $\chi \chi \to V \to d_i \bar{d}_j$.
As discussed in Section~\ref{sec:loopdecays}, the loop contributions
are suppressed by the mass of the light up-type quark that the mediator couples to.
In Section~\ref{sec:DMheavy}, we have shown
that in the region where two-body final states are allowed, the relic abundance requires the couplings to be fairly large, therefore in the light mass region 
the $\chi \chi$ annihilation rate is, without doubts, too slow for the stable particle $\chi$ not to overpopulate the Universe.

One possible way-out is to turn on the $a_L$ coupling. In this case, we open back a two body decay $\chi \chi \to b \bar{d}$ and $\bar{b} d$, and similar numerical results as in Section~\ref{sec:DMheavy} at the price of a less minimal scenario. A second possibility is to have a very small $a_R$ so that $V$ is long lived at the price of suppressing the monotop signal beyond any hope of detectability, or complicating the dark sector so that $\chi$ is not the dark matter candidate. 

Following this argument, we can state that the minimal monotop scenario is excluded by dark matter relic abundance constraints when the mediator is lighter that the top.

\section{Conclusions}
\label{sec:conclusions}

Monotop final states comprised of a single top quark produced in association with missing
energy can be a striking sign of new physics at the LHC.
The main production mechanisms can be divided into two classes: resonant production, where a
heavy coloured boson is first produced in the $s$-channel and further decays via its
couplings to a single top quark and an invisible neutral fermion, and non-resonant production
where the top quark is produced in association with an invisible boson that couples to top and
up (or charm) quarks.
A complete and model independent parametrisation of the two channels has been provided in
Ref.~\cite{Andrea:2011ws}.
In the present work, we have revisited this description by embedding the effective interactions
in an SU(2)$_L \times$ U(1)$_Y$ invariant formalism.
In doing so, we have shown that, depending on the chirality of the tops, a complete model contains necessarily extra states and couplings that may spoil the monotop signal, or add more new physics signatures that should be studied in association with the monotop one.

We have identified two minimal setups. In the first case,
a scalar field is resonantly produced by the fusion of a pair of down-type quarks and
couples to a right-handed top quark and a new
invisible fermion, like a right-handed stop in $R$-parity violating supersymmetry. In the second case, a vector
state couples to right-handed top and up quarks and decays dominantly into new invisible fields,
like in models of dark matter where the dark sector couples to the Standard Model via a
flavour-sensitive mediator.
We have further investigated the phenomenology of the second class of models that can be split
into two subclasses, depending on the mass of the mediator.

For mediators lighter than the top quark, their visible decay modes are either loop-suppressed
or phase-space-suppressed, or both. Nevertheless, one always needs to add (and tune the couplings
of) an invisible
field to prevent the mediator from decaying inside a typical hadron collider detector as
this would otherwise spoil the monotop signature originally motivating the model.
An important feature of these scenarios is that they allow for the top quark to decay into the
mediator and an extra jet. This feature can enhance the monotop production rate, as the monotop
system can be produced in association with an extra jet from
$t \bar{t}$ events when one of the top quarks decays in the exotic channel. Such events
could also be searched for in standard typical single-top searches, as they are
expected to populate signal regions of associated analyses.
We have indeed observed that a CMS analysis of single top events could imply significant
constraints on the mediator couplings, competitive and sometimes
stronger than those obtained from monotop searches.

Scenarios with a mediator mass above the top threshold have a very different phenomenology
as the mediator decays significantly into top quarks and jets. One needs a large coupling
to the invisible sector in order to preserve the monotop signature.
Describing the dark sector with a new fermion $\chi$, we have found that the latter could be
a viable dark matter candidate if heavier than half the top quark mass, with a correct relic abundance
driven by its annihilation via an $s$-channel mediator into a top and an up quark.
We have used relic abundance constraints to derive lower bounds on the product of the couplings
of the mediator to quarks and to the dark matter candidate.
We have then further restricted the monotop parameter space by combining cosmological
and collider results and enforcing the mediator to decay mostly invisibly.
We have found that the issue of the perturbativity of the model could be raised
for dark matter masses close to the top mass and that the parameter space
turns out to be largely constrained when the $\chi$ fermion is demanded to reproduce the
observed relic density. However, a large portion of the parameter space is still
left unconstrained by current data and future experimental results are in order,
in particular analyzing a same-sign top quark pair final state arising from the visible decays of the mediator.

\acknowledgments
We thank T.~Theveneaux-Pelzer, J.~Donini and F.~Maltoni for useful discussions and comments. We acknowledge 
support by the Theory-LHC France initiative of CNRS/IN2P3.
G.C. and A.D. acknowledge partial support from the Labex-LIO (Lyon Institute of Origins) under grant ANR-10-LABX-66 
and FRAMA (FR3127, F\'ed\'eration de Recherche ``Andr\'e Marie Amp\`ere"). A.D. is partially supported by Institut 
Universitaire de France.

\bibliographystyle{JHEP}
\bibliography{bibliography}

\providecommand{\href}[2]{#2}\begingroup\raggedright\begin{thebibliography}{10}

\bibitem{atlas:susytwiki}
 {\em
  https://twiki.cern.ch/twiki/bin/view/AtlasPublic/SupersymmetryPublicResults}.

\bibitem{cms:susytwiki}
 {\em https://twiki.cern.ch/twiki/bin/view/CMSPublic/PhysicsResultsSUS}.

\bibitem{Berger:1999zt}
E.~L. Berger, B.~Harris, and Z.~Sullivan, {\it {Single top squark production
  via R-parity violating supersymmetric couplings in hadron collisions}},  {\em
  Phys.Rev.Lett.} {\bf 83} (1999) 4472--4475,
  [\href{http://xxx.lanl.gov/abs/hep-ph/9903549}{{\tt hep-ph/9903549}}].

\bibitem{delAguila:1999ac}
F.~del Aguila, J.~Aguilar-Saavedra, and L.~Ametller, {\it {Z t and gamma t
  production via top flavor changing neutral couplings at the Fermilab
  Tevatron}},  {\em Phys.Lett.} {\bf B462} (1999) 310--318,
  [\href{http://xxx.lanl.gov/abs/hep-ph/9906462}{{\tt hep-ph/9906462}}].

\bibitem{Berger:2000zk}
E.~L. Berger, B.~Harris, and Z.~Sullivan, {\it {Direct probes of R-parity
  violating supersymmetric couplings via single top squark production}},  {\em
  Phys.Rev.} {\bf D63} (2001) 115001,
  [\href{http://xxx.lanl.gov/abs/hep-ph/0012184}{{\tt hep-ph/0012184}}].

\bibitem{Desai:2010sq}
N.~Desai and B.~Mukhopadhyaya, {\it {R-parity violating resonant stop
  production at the Large Hadron Collider}},  {\em JHEP} {\bf 1010} (2010) 060,
  [\href{http://xxx.lanl.gov/abs/1002.2339}{{\tt 1002.2339}}].

\bibitem{Davoudiasl:2011fj}
H.~Davoudiasl, D.~E. Morrissey, K.~Sigurdson, and S.~Tulin, {\it {Baryon
  Destruction by Asymmetric Dark Matter}},  {\em Phys.Rev.} {\bf D84} (2011)
  096008, [\href{http://xxx.lanl.gov/abs/1106.4320}{{\tt 1106.4320}}].

\bibitem{Andrea:2011ws}
J.~Andrea, B.~Fuks, and F.~Maltoni, {\it {Monotops at the LHC}},  {\em
  Phys.Rev.} {\bf D84} (2011) 074025,
  [\href{http://xxx.lanl.gov/abs/1106.6199}{{\tt 1106.6199}}].

\bibitem{Kamenik:2011nb}
J.~F. Kamenik and J.~Zupan, {\it {Discovering Dark Matter Through Flavor
  Violation at the LHC}},  {\em Phys.Rev.} {\bf D84} (2011) 111502,
  [\href{http://xxx.lanl.gov/abs/1107.0623}{{\tt 1107.0623}}].

\bibitem{Dong:2011rh}
Z.~Dong, G.~Durieux, J.-M. Gerard, T.~Han, and F.~Maltoni, {\it {Baryon number
  violation at the LHC: the top option}},  {\em Phys.Rev.} {\bf D85} (2012)
  016006, [\href{http://xxx.lanl.gov/abs/1107.3805}{{\tt 1107.3805}}].

\bibitem{Wang:2011uxa}
J.~Wang, C.~S. Li, D.~Y. Shao, and H.~Zhang, {\it {Search for the signal of
  monotop production at the early LHC}},  {\em Phys.Rev.} {\bf D86} (2012)
  034008, [\href{http://xxx.lanl.gov/abs/1109.5963}{{\tt 1109.5963}}].

\bibitem{Fuks:2012im}
B.~Fuks, {\it {Beyond the Minimal Supersymmetric Standard Model: from theory to
  phenomenology}},  {\em Int.J.Mod.Phys.} {\bf A27} (2012) 1230007,
  [\href{http://xxx.lanl.gov/abs/1202.4769}{{\tt 1202.4769}}].

\bibitem{Kumar:2013jgb}
A.~Kumar, J.~N. Ng, A.~Spray, and P.~T. Winslow, {\it {Tracking Down the Top
  Quark Forward-Backward Asymmetry with Monotops}},  {\em Phys.Rev.} {\bf D88}
  (2013) 075012, [\href{http://xxx.lanl.gov/abs/1308.3712}{{\tt 1308.3712}}].

\bibitem{Alvarez:2013jqa}
E.~Alvarez, E.~C. Leskow, J.~Drobnak, and J.~F. Kamenik, {\it {Leptonic
  Monotops at LHC}},  {\em Phys.Rev.} {\bf D89} (2014) 014016,
  [\href{http://xxx.lanl.gov/abs/1310.7600}{{\tt 1310.7600}}].

\bibitem{Agram:2013wda}
J.-L. Agram, J.~Andrea, M.~Buttignol, E.~Conte, and B.~Fuks, {\it {Monotop
  phenomenology at the Large Hadron Collider}},  {\em Phys.Rev.} {\bf D89}
  (2014) 014028, [\href{http://xxx.lanl.gov/abs/1311.6478}{{\tt 1311.6478}}].

\bibitem{Fuks:2014uka}
B.~Fuks, J.~Proudom, J.~Rojo, and I.~Schienbein, {\it {Characterizing New
  Physics with Polarized Beams at High-Energy Hadron Colliders}},  {\em JHEP}
  {\bf 1405} (2014) 045, [\href{http://xxx.lanl.gov/abs/1403.2383}{{\tt
  1403.2383}}].

\bibitem{Ng:2014pqa}
J.~N. Ng and A.~de~la Puente, {\it {Probing Radiative Neutrino Mass Generation
  through Monotop Production}},  \href{http://xxx.lanl.gov/abs/1404.1415}{{\tt
  1404.1415}}.

\bibitem{Aaltonen:2012ek}
{\bf CDF} Collaboration, T.~Aaltonen {\em et.~al.}, {\it {Search for a dark
  matter candidate produced in association with a single top quark in
  $p\bar{p}$ collisions at $\sqrt{s} = 1.96$ TeV}},  {\em Phys.Rev.Lett.} {\bf
  108} (2012) 201802, [\href{http://xxx.lanl.gov/abs/1202.5653}{{\tt
  1202.5653}}].

\bibitem{CMS:2014hba}
{\bf CMS} Collaboration, {\it {Search for new physics with monotop final states
  in pp collisions at sqrt(s) = 8 TeV}},
  \href{http://xxx.lanl.gov/abs/CMS-PAS-B2G-12-022}{{\tt CMS-PAS-B2G-12-022}}.

\bibitem{Khachatryan:2014uma}
{\bf CMS} Collaboration, V.~Khachatryan {\em et.~al.}, {\it {Search for monotop
  signatures in proton-proton collisions at $\sqrt{s}$ = 8 TeV}},
  \href{http://xxx.lanl.gov/abs/1410.1149}{{\tt 1410.1149}}.

\bibitem{Aad:2014wza}
{\bf ATLAS} Collaboration, G.~Aad {\em et.~al.}, {\it {Search for invisible
  particles produced in association with single-top-quarks in proton-proton
  collisions at $\sqrt{s}$ = 8 TeV with the ATLAS detector}},
  \href{http://xxx.lanl.gov/abs/1410.5404}{{\tt 1410.5404}}.

\bibitem{Davoudiasl:2010am}
H.~Davoudiasl, D.~E. Morrissey, K.~Sigurdson, and S.~Tulin, {\it {Hylogenesis:
  A Unified Origin for Baryonic Visible Matter and Antibaryonic Dark Matter}},
  {\em Phys.Rev.Lett.} {\bf 105} (2010) 211304,
  [\href{http://xxx.lanl.gov/abs/1008.2399}{{\tt 1008.2399}}].

\bibitem{Jack:1982sr}
I.~Jack and H.~Osborn, {\it {General Two Loop Beta Functions for Gauge Theories
  With Arbitrary Scalar Fields}},  {\em J.Phys.} {\bf A16} (1983) 1101.

\bibitem{Li:1992dt}
L.-F. Li, {\it {Oblique electroweak corrections from heavy scalar fields}},
  {\em Z.Phys.} {\bf C58} (1993) 519--522.

\bibitem{Bhattacharyya:1993zy}
G.~Bhattacharyya, A.~Kundu, T.~De, and B.~Dutta-Roy, {\it {Effects of
  isodoublet color - octet scalar bosons on oblique electroweak parameters}},
  {\em J.Phys.} {\bf G21} (1995) 153.

\bibitem{Aad:2013pqd}
{\bf ATLAS} Collaboration, G.~Aad {\em et.~al.}, {\it {Search for long-lived,
  multi-charged particles in pp collisions at $\sqrt{s}$=7 TeV using the ATLAS
  detector}},  {\em Phys.Lett.} {\bf B722} (2013) 305--323,
  [\href{http://xxx.lanl.gov/abs/1301.5272}{{\tt 1301.5272}}].

\bibitem{Chatrchyan:2013oca}
{\bf CMS} Collaboration, S.~Chatrchyan {\em et.~al.}, {\it {Searches for
  long-lived charged particles in pp collisions at $\sqrt{s}$=7 and 8 TeV}},
  {\em JHEP} {\bf 1307} (2013) 122,
  [\href{http://xxx.lanl.gov/abs/1305.0491}{{\tt 1305.0491}}].

\bibitem{Alwall:2014bza}
J.~Alwall, C.~Duhr, B.~Fuks, O.~Mattelaer, D.~G. Ozturk, {\em et.~al.}, {\it
  {Computing decay rates for new physics theories with FeynRules and
  MadGraph5/aMC@NLO}},  \href{http://xxx.lanl.gov/abs/1402.1178}{{\tt
  1402.1178}}.

\bibitem{Belyaev:2012qa}
A.~Belyaev, N.~D. Christensen, and A.~Pukhov, {\it {CalcHEP 3.4 for collider
  physics within and beyond the Standard Model}},  {\em Comput.Phys.Commun.}
  {\bf 184} (2013) 1729--1769, [\href{http://xxx.lanl.gov/abs/1207.6082}{{\tt
  1207.6082}}].

\bibitem{Kolb:1990vq}
E.~W. Kolb and M.~S. Turner, {\em The Early Universe}.
\newblock Addison-Wesley, 1990.
\newblock Frontiers in Physics, 69.

\bibitem{Aaltonen:2013kna}
{\bf CDF} Collaboration, T.~A. Aaltonen {\em et.~al.}, {\it {Direct Measurement
  of the Total Decay Width of the Top Quark}},  {\em Phys.Rev.Lett.} {\bf 111}
  (2013), no.~20 202001, [\href{http://xxx.lanl.gov/abs/1308.4050}{{\tt
  1308.4050}}].

\bibitem{Beringer:1900zz}
{\bf Particle Data Group} Collaboration, J.~Beringer {\em et.~al.}, {\it
  {Review of Particle Physics (RPP)}},  {\em Phys.Rev.} {\bf D86} (2012)
  010001.

\bibitem{Khachatryan:2014iya}
{\bf CMS} Collaboration, V.~Khachatryan {\em et.~al.}, {\it {Measurement of the
  t-channel single-top-quark production cross section and of the $\mid V_{tb}
  \mid$ CKM matrix element in pp collisions at $\sqrt{s}$= 8 TeV}},  {\em JHEP}
  {\bf 1406} (2014) 090, [\href{http://xxx.lanl.gov/abs/1403.7366}{{\tt
  1403.7366}}].

\bibitem{Aad:2012ux}
{\bf ATLAS} Collaboration, G.~Aad {\em et.~al.}, {\it {Measurement of the
  $t$-channel single top-quark production cross section in $pp$ collisions at
  $\sqrt{s}=7$ TeV with the ATLAS detector}},  {\em Phys.Lett.} {\bf B717}
  (2012) 330--350, [\href{http://xxx.lanl.gov/abs/1205.3130}{{\tt 1205.3130}}].

\bibitem{Kraml:2012sg}
S.~Kraml, B.~Allanach, M.~Mangano, H.~Prosper, S.~Sekmen, {\em et.~al.}, {\it
  {Searches for New Physics: Les Houches Recommendations for the Presentation
  of LHC Results}},  {\em Eur.Phys.J.} {\bf C72} (2012) 1976,
  [\href{http://xxx.lanl.gov/abs/1203.2489}{{\tt 1203.2489}}].

\bibitem{Christensen:2008py}
N.~D. Christensen and C.~Duhr, {\it {FeynRules - Feynman rules made easy}},
  {\em Comput.Phys.Commun.} {\bf 180} (2009) 1614--1641,
  [\href{http://xxx.lanl.gov/abs/0806.4194}{{\tt 0806.4194}}].

\bibitem{Alloul:2013bka}
A.~Alloul, N.~D. Christensen, C.~Degrande, C.~Duhr, and B.~Fuks, {\it
  {FeynRules 2.0 - A complete toolbox for tree-level phenomenology}},  {\em
  Comput.Phys.Commun.} {\bf 185} (2014) 2250--2300,
  [\href{http://xxx.lanl.gov/abs/1310.1921}{{\tt 1310.1921}}].

\bibitem{Degrande:2011ua}
C.~Degrande, C.~Duhr, B.~Fuks, D.~Grellscheid, O.~Mattelaer, {\em et.~al.},
  {\it {UFO - The Universal FeynRules Output}},  {\em Comput.Phys.Commun.} {\bf
  183} (2012) 1201--1214, [\href{http://xxx.lanl.gov/abs/1108.2040}{{\tt
  1108.2040}}].

\bibitem{Alwall:2014hca}
J.~Alwall, R.~Frederix, S.~Frixione, V.~Hirschi, F.~Maltoni, {\em et.~al.},
  {\it {The automated computation of tree-level and next-to-leading order
  differential cross sections, and their matching to parton shower
  simulations}},  {\em JHEP} {\bf 1407} (2014) 079,
  [\href{http://xxx.lanl.gov/abs/1405.0301}{{\tt 1405.0301}}].

\bibitem{Sjostrand:2006za}
T.~Sjostrand, S.~Mrenna, and P.~Z. Skands, {\it {PYTHIA 6.4 Physics and
  Manual}},  {\em JHEP} {\bf 0605} (2006) 026,
  [\href{http://xxx.lanl.gov/abs/hep-ph/0603175}{{\tt hep-ph/0603175}}].

\bibitem{deFavereau:2013fsa}
{\bf DELPHES 3} Collaboration, J.~de~Favereau {\em et.~al.}, {\it {DELPHES 3, A
  modular framework for fast simulation of a generic collider experiment}},
  {\em JHEP} {\bf 1402} (2014) 057,
  [\href{http://xxx.lanl.gov/abs/1307.6346}{{\tt 1307.6346}}].

\bibitem{Dumont:2014tja}
B.~Dumont, B.~Fuks, S.~Kraml, S.~Bein, G.~Chalons, {\em et.~al.}, {\it {Towards
  a public analysis database for LHC new physics searches using MadAnalysis
  5}},  \href{http://xxx.lanl.gov/abs/1407.3278}{{\tt 1407.3278}}.

\bibitem{Conte:2012fm}
E.~Conte, B.~Fuks, and G.~Serret, {\it {MadAnalysis 5, A User-Friendly
  Framework for Collider Phenomenology}},  {\em Comput.Phys.Commun.} {\bf 184}
  (2013) 222--256, [\href{http://xxx.lanl.gov/abs/1206.1599}{{\tt 1206.1599}}].

\bibitem{Conte:2014zja}
E.~Conte, B.~Dumont, B.~Fuks, and C.~Wymant, {\it {Designing and recasting LHC
  analyses with MadAnalysis 5}},  {\em Eur.Phys.J.} {\bf C74} (2014) 3103,
  [\href{http://xxx.lanl.gov/abs/1405.3982}{{\tt 1405.3982}}].

\end{thebibliography}\endgroup

\end{document}